# A New BAT and PageRank algorithm for Propagation Probability in Social Networks

Wei-Chang Yeh, Chia-Ling Huang, Tzu-Yun Hsu, Zhenyao Liu, Shi-Yi Tan

[1] Department of Industrial Engineering and Engineering Management, National Tsing Hua University, Taiwan

**Abstract**

Social networks have increasingly become important and popular in modern times. Moreover, the influence of social networks plays a vital role in various organizations including government organizations, academic research or corporate organizations. Therefore, how to strategize the optimal propagation strategy in social networks has also become more important. By increasing the precision of evaluating the propagation probability of social network, it can indirectly influence the investment of cost, manpower and time for information propagation to achieve the best return.

This study proposes a new algorithm, which includes a scale-free network, Barabási-Albert model, Binary-Addition-Tree (BAT) algorithm, PageRank algorithm, personalized PageRank algorithm and a new BAT algorithm, to calculate the propagation probability in social networks.

The results obtained after implementing the simulation experiment of social network models show the studied model and the proposed algorithm provide an effective method to increase the efficiency of information propagation in social networks. In this way, the maximum propagation efficiency is achieved with the minimum investment.

## 1. Introduction

With the commercial effect of viral marketing and word-of mouth marketing, social networks are regarded as the medium for propagating information, ideas and influence among nodes [1-4]. The propagated influence on information propagation in social networks, which use the behavior of user information propagation in social networks to evaluate the propagated influence and find out the key nodes in order to have the greatest influence on social networks during the propagation of these nodes [1, 5], is hot research area [1, 5-9]. In other words, investors focus on the high connectivity and high propagation of social networks and accurately invest information in the source nodes of propagation rather than invest a lot of resources in many nodes. The use of social network propagation achieves the purpose of information propagation, which is also the advantage of social networks [10].

In fact, people's lives and sources of information are largely affected by social networks [11] so that understand how information propagates on social networks is crucial for designing effective promotion strategies and preventing the propagation of malicious information. It is found the social network is a scale-free network in the real-world [12, 13], and the degree distribution of the scale-free network follows the power law distribution [14]. Many practical and important networks are scale-free networks, which share information between nodes in the network through network connections, such as the World Wide Web, power grids, etc. [14, 15].

Many researches have been devoted to modeling the propagation process in social networks [16-20]. Among the different models of scale-free networks, the Barabási-Albert model proposed by Barabási and Albert is more common and has many applications [15]. Therefore, this study adopts the Barabási- Albert model to simulate the propagation in social networks and generate test data to verify the performance of the proposed algorithm.

Measuring the probability of propagation in social networks will vary depending on the focus of research. The most intuitive way is to use the number of node connections that the probability of page appearance is high when the number of node connections is large, which is called In-degree [21, 22]. However, using this method will lead to a large number of spam links generated by intentional people in order to increase the probability of the page. The TwitterRank algorithm assigns different propagation weights for Twitter users to track the similarity of different specific topics and web page structures, while the probability will vary with the posting rate of the post [23]. The PageRank algorithm is based on web page links, its PageRank value represents random clicks on the page as the probability distribution of reaching any particular node, and the source of its value is based on the PageRank values of all other pages linked to it [24]. This study adopts the PageRank algorithm originally used by Google to analyze the relevance and importance of web pages, and uses the PageRank value to measure the importance of nodes to convert it into an appropriate propagation probability. However, studies in [25-27] pointed out that users of social networks have different topic preferences resulting in different topics with different propagation effects. Therefore, in order to analyze the influence of different user preferences on the propagation probability, this study also uses Personalized PageRank to measure the different node weights due to different node themes [28-30].

There are many methods to search for the propagation path of social networks. Clauset et al. proposed the hierarchical structure can be inferred from the network and applied to solve the path prediction problem in social networks [31]. Backstrom and Leskovec proposed a prediction algorithm based on supervised random propagation path, which the random path is more convincing because the transfer probability between nodes is different. However, it has not been applied in practice because the model requires a lot of computing time [32]. Whether based on random paths or fully supervised, these methods are generally complex and difficult to understand, and are more suitable for problems with a fixed number of sink nodes. However, the propagation process in social networks studied in this work does not have a fixed sink node so that the propagation problem in social networks to be predicted in this study requires an easy-to-understand new method to predict its propagation path and propagation probability.

This study adopts Binary-Addition Tree algorithm (BAT) to predict the propagation in social networks [33]. BAT algorithm is similar to Depth-First-Search (DFS) [34] and Breadth-First-Search (BFS) [35]. However, BAT algorithm in addition to being easy to understand and program, it has the main advantage of flexibly searching all propagation states in the binary network, which makes the path exhaustion simpler and easier to change for different applications [33].

This study aims to propose a new method to calculate the probability of propagation area in social networks and to quickly identify nodes in the network that require close attention. In addition to adopting the traditional PageRank algorithm and the personalized PageRank algorithm, this study also proposes a novel state concept and a new method based on the combination of BAT algorithm and PageRank algorithm to predict the area and probability of possible impact of propagation in the modeled social network, and perform Social Network Analysis (SNA).

The remained structure of this study is as follows. Section 2 introduces social network and its propagation model, scale-free network, Barabási- Albert model, PageRank algorithm, personalized PageRank algorithm and BAT algorithm. Section 3 shows the research method and proposes the propagation status in social networks, demonstrates how to present the propagation path numerically the calculation of the propagation probability, and combines the BAT algorithm with the PageRank algorithm to propose a new BAT algorithm for this study. The propagation in social networks is simulated and the experimental results are presented to assist the analysis in Section 4. Finally, we summarize the research contributions and point out future work in Section 5.

## 2. Preliminaries

This study uses the Barabási-Albert model to model the scale-free problem of propagation probability in social networks, and adopts the modified BAT algorithm combined with the PageRank algorithm to make predictions. Before proceeding to research methods, this section introduces the background of social networks, scale-free networks, Barabási-Albert model, PageRank algorithm and BAT algorithm.

### 2.1 Social networks

The term of social network was proposed by Barnes in 1954 [20], which is used to describe a community structure of the relationship between members in a complex social system and can be presented graphically. Theoretically, a social network should contain two elements: nodes and edges (connection relationships) [19], the former can represent organizations, communities or individuals, etc., and the latter can represent the relationship of friendships, business partners, etc. In this study, all web pages on the Internet are regarded as nodes such as social platforms or fan pages, and hyperlinks are regarded as edges [36].

Due to the availability of large datasets and multi-domain applications, social networks have been rapidly developed and applied in many fields in the past decade [37-42]. The customer data of social network is used to discover social network groups and develop a comprehensive algorithm to enhance the model of the advertising system [37]. Travelers use online social networks to maintain personal-related information, and back-end systems analyze and record the traveler's location to provide the best itinerary according to the route and traveler's preferences [38]. By analyzing the social behavior of social structure nodes, the social attributes of the social network are mined to provide humanized services [39]. Considering the social network in the driving environment, the structure of the Video Sensor Node (VSN) and communication architecture is described to simulate the similar goals of vehicles, passengers and drivers on the road in the virtual community commuters, interests or action patterns to contribute to future intelligent traffic control [40].

A large number of studies have found that the real social network is a network, which appears to be random, actually follows the degree distribution of nodes presents a power distribution [43-45]. In research [11], a propagation tree of 227 actual blog page nodes is presented. The propagation tree takes the source node as the center point. Nodes with less than four propagation layers have an average branch count greater than 1, while the number of branches for nodes with higher propagation layers is almost 1, as shown in Figure 1.

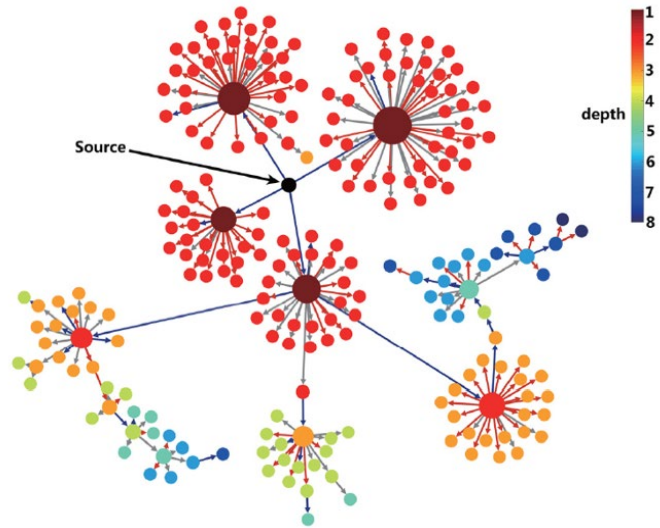

Figure 1. 227 blog page nodes propagation tree and propagation layers

Coincidentally, in order to present the back-end computing system of its own platform and prove that it has 2.2 billion monthly active users, Facebook presents all Facebook social platform users with blue dots on the black layer and connects the relationship between users with blue lines, which is drawn in Figure 2. The clever thing is that the black layer gradually takes on the shape of the world map. Through Figure 2, we can know the actual communication characteristics of the social network.

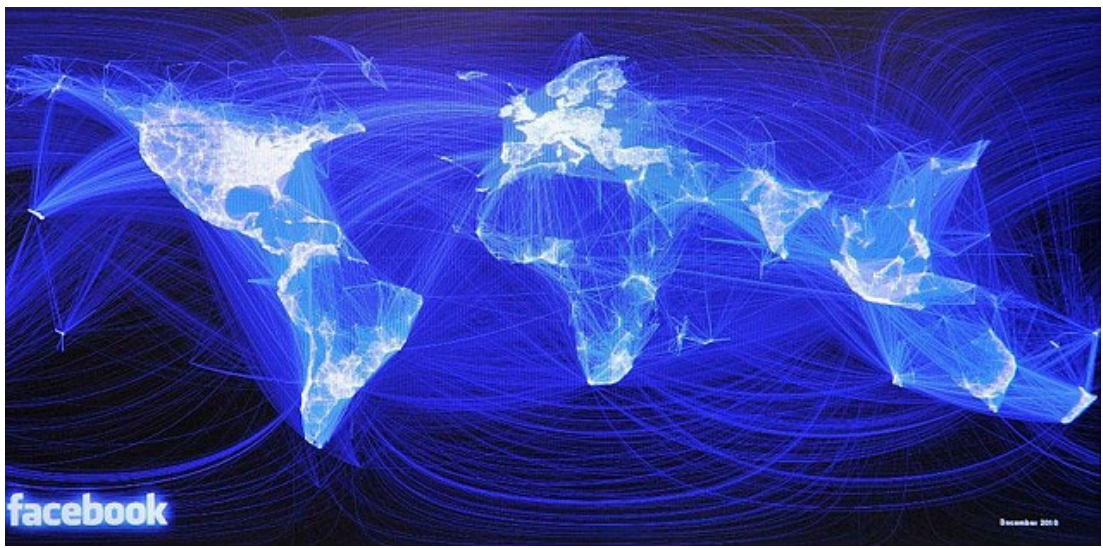

Figure 2. Facebook depicts all connected community users

In the branch structure of nodes of the propagation tree in Figures 1-2, it is found that the propagation path of the social network is centered on the source node. Most nodes in the network are only connected to a few key nodes. Relatively speaking, there are very few key nodes connected to more nodes in the network. The network of this feature type is the scale-free network proposed by Barabási in 1999 [15], and a few key nodes are called hub.

Many important and huge networks are scale-free networks, such as the World Wide Web (WWW) [15], the internet [46], and the metabolic network [47], which the degree distribution of the network follows the power distribution. The social network is one of the applications of the WWW [48], and many studies have proved that the social network is a scale-free network [15, 49-51].

The degree distribution of scale-free network is a power distribution [15] and most nodes are only connected to the hub node so that the hub node can master the overall propagation of the social network [52]. Moreno and Vazquez, in their research on whether the proportion of immune nodes on scale-free network can eradicate epidemic viruses, believe that would be better than randomly picking nodes across the network to make them immune if the hub nodes on scale-free network have immune effects against epidemics [44]. Therefore, mastering the hub nodes is also very important in the prediction of propagation in social networks.

**2.2 Scale-free networks and Barabási-Albert model**

As mentioned in Section 2.1, many scholars have proved that the social network is a scale-free network. In the scale-free network, when a node has $k$ connections, their degree distribution follows a power distribution [15], such as Eq. (1).

$$P(k) \sim k^{-\eta} \tag{1}$$

The power distribution has a higher incidence when the number of $k$ is small, and the number of occurrences is very rare when its number grows [13, 51]. In the power distribution function, even if the value of $k$ only increases a little, the effect on $P(k)$ can be affected by the power series, which the power is negative, that is $k^{-\eta}$ as shown in Figure 3.

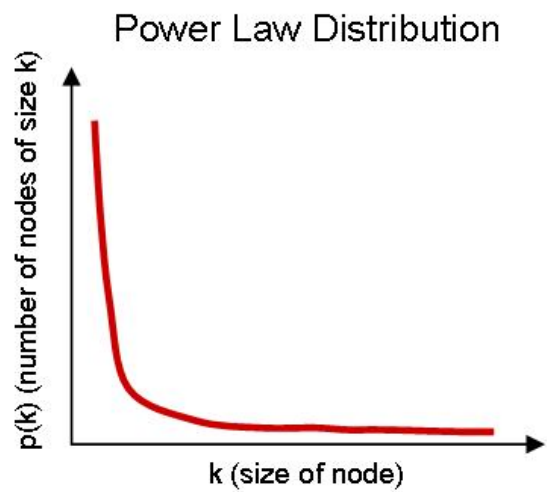

Figure 3. Power distribution plot

The scale-free web follows a power distribution was discovered by Barabási and Albert on their collaborative study of the WWW in 1998. The WWW consists of several highly connected pages, most of which have no more than 4 hyperlinks but very few pages have thousands of links [15]. The social networks in real world will grow in size by adding new nodes, which are starting from a small source node. By adding new nodes, the number of nodes increases regularly throughout the network [13-15, 53-54]. For example, the WWW grows exponentially with the addition of new pages and research citations grow with the publication of new papers, which Barabási and Albert call the scale-free network [15].

The network in real world, the newly added nodes are generally connected to the nodes with more connections in the original network that this phenomenon is called rich-get-richer [54-57]. For example, new webpages tend to link to already popular webpages because such webpages are easy to find and widely known, such as the homepage of the website. The generation of new nodes in the network and the preferential connection of nodes with more connections in the original network prompted Barabási and Albert to propose the Barabási-Albert model in 1999 [15].

The algorithm of Barabási-Albert model is based on two modes [55, 57], as follows:

1. Growth mode: The network in real world continues to expand and grow, such as adding new friends on social platforms or creating new pages on the Internet.

2. Preferential connection mode: New nodes tend to connect to nodes with more connections in the original network when they join. When selecting a new node to connect, it is assumed that there are *n* nodes, where *n*=0, 1, 2, …, (*n*−1), the probability of the new node connecting to node *i*, namely $P_i$, depends on the degree of node *i*, namely $k_i$, as in Eq.(2).

$$P_i = \frac{k_i}{\sum_{j=0}^{n-1} k_j} \quad (2)$$

To understand how scale-free networks propagate, Barabási and Albert presented a simple process for generating scale-free networks in Figure 4 [53]. Starting from three connected nodes (*t*=1), the original nodes in the layer are represented by solid lines, and the newly added nodes are represented by hollow circles.

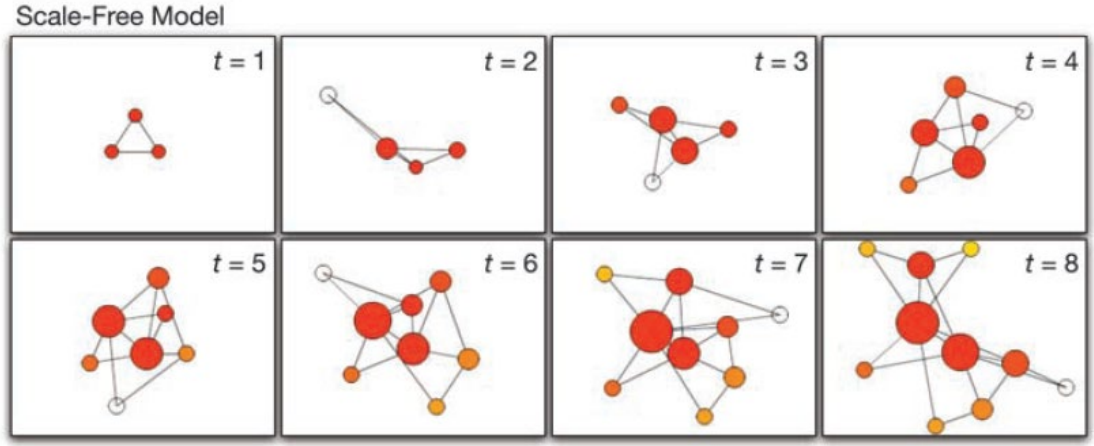

Figure 4. Scale-free network generation process

Due to the growth mode and priority connection mode, the newly added nodes prefer to connect to the original nodes with more connections leading to the natural output of several highly connected nodes, which means that the probability of selecting a new node is proportional to the number of node connections.

A typical feature of scale-free network is that the node degree obeys a power distribution, most nodes in the network are only connected to a few nodes and very few nodes are connected to most nodes such key nodes are called hubs. The existence of such nodes makes the scale-free network more resistant to sudden failures but it is also vulnerable to collective attacks [15], as shown in Figure 5. From Eq. (2), it can be seen that most nodes are non-hub nodes and the degree of connection $k_i$ is very small in the scale-free network. Based on the preferential connection mode, some highly numbered non-hub nodes can be gradually transformed into hobs as new nodes are more likely to be connected to it [15, 53-37]. In other words, all nodes approximately follow a power distribution [14, 50].

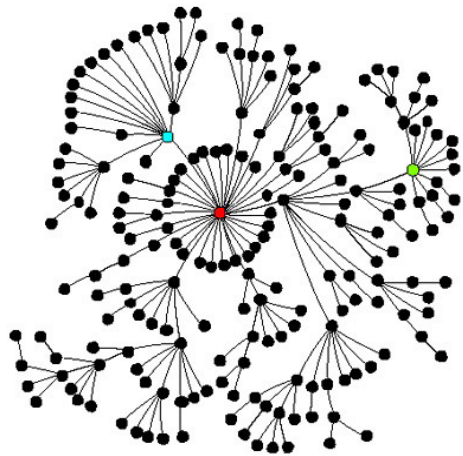

Figure 5. Scale-free network

Among different models of scale-free networks, the Barabási-Albert model proposed by Barabási and Albert is more common and has many applications for scale-free networks [15]. Therefore, this study uses the Barabási-Albert model to construct a scale-free network to predict the propagation probability of social networks.

**2.3 PageRank algorithm**

The PageRank algorithm is an algorithm previously invented by Google Inc. to rank the importance of pages in search results by their company's search engine [58]. The PageRank value refers to the possibility of a web page being seen. Each web page has an individual PageRank value, which depends on the link relationship between the web pages. A higher PageRank value means that a web page is more popular [22, 24, 58, 59]

**2.3.1 PageRank**

The PageRank algorithm considers the connection structure of network to measure the probability, and its complete definition is as Eq. (3) [24]. Eq. (3) implements the idea of the PageRank value, that is, the PageRank value of each web page is positively correlated with the PageRank value of web pages linked to the web page [24, 58]. Hence, the more pages that are linked to, the more PageRank values can be accumulated. And a website with higher PageRank value can obtain a larger PageRank value. The calculation process is shown in Table 1 [60].

$$PR(u) = (1-d) + d \sum_{v \in B(u)} \frac{PR(v)}{N_v} \qquad (3)$$

where, $u$: web page. $B(u)$: a set of pages connecting $u$. $PR(u)$: the PageRank value of page $u$. $PR(v)$: the PageRank value of page $v$. $N_v$: the number of outgoing links from page $v$. $d$: a damping factor between 0 and 1, with a default value of 0.85. When $d = 0.85$, there is an 85% probability that Internet users will continue to the next page after visiting a particular page.

Table 1. PageRank algorithm

| | |
|---|---|
| Input: | $G(V, E)$, where $V = \{0, 1, 2, …, u–1\}$. |
| Output: | PageRank value of nodes. |
| Step P0： | Let $t = 0$, set the initial PageRank value of all nodes to $1/u$. |
| Step P1： | According to Eq. (3), update the PageRank values of all nodes. |
| Step P2： | For nodes with $N_v = 0$, evenly distribute the PageRank values of |

other nodes.

Step P3： If the PageRank value of all nodes does not change, stop, otherwise let $t = t +1$ and go to P0.

The PageRank algorithm iteratively updates a node's PageRank value by accumulating the PageRank value of each node connected to it, then dividing by the number of the referencing node. First, initialize all PageRank values with the same weight, then update the PageRank values of all nodes with Eq. (3). The $N_v$ = 0 in step P2 means that the number of outgoing links from page $v$ is 0, such as going to the page directly through the URL. At this time, the PageRank value of other nodes is enenly distributed. The update is repeated until the PageRank value converges to a unique constant.

**2.3.2 Personalized PageRank**

The PageRank algorithm introduces the concept of random viewers into the calculation process, which gives equal weight to all pages of incoming links. In other words, the PageRank algorithm only uses the link structure of network and cannot judge the similarity in the content of web pages. In addition, the PageRank algorithm distributes the weights evenly according to the number of outgoing links so that the pages that are not related to the theme get the same attention as the pages that are related to the theme resulting in a theme shift [61]. No matter how good the PageRank value is, if the page is off topic, the PageRank value returned to the user has little value [30, 61, 62].

In order to solve the problem of topic shift in PageRank algorithm, Page et al. proposed a personalized PageRank method to measure the different importance of nodes [63]. Personalized PageRank affects the assigned weight of each node in the random walk and uses the Personalized vector, namely $r_u$, to bias the random walk process to a specific node [28-30, 62, 64], as shown in Eq. (4).

$$PR(u) = (1-d)r_u + d \sum_{v \in B(u)} \frac{PR(v)}{N_v} \qquad (4)$$

where, $u$: web page. $B(u)$: a set of pages connecting $u$. $PR(u)$: the PageRank value of page $u$. $PR(v)$: the PageRank value of page $v$. $N_v$: the number of outgoing links from page $v$. $d$: a damping factor between 0 and 1. $i$: the preferred target page, when $u=i$, $r_u$ =1; otherwise, $r_u$=0.

From Eq. (3) and Eq. (4), it can be found that the difference between personalized PageRank and PageRank is that the personalized vector in PageRank moves to any random node with the same probability; while personalized PageRank adjusts the personalized vector so that random viewers always move to a node or set of nodes that we are interested in, rather than from all nodes in the network to any node. If we want random viewers to move to a specific node such as page $i$, the personalized vector is 1, that is $r_u$=1, and the personalized vectors of the rest of nodes are assigned 0 [28-30].

The personalized PageRank can send random viewers to nodes related to preferences according to user-specified preferences, change node weights [29], and improve the problem of PageRank and topic bias. This study uses personalized PageRank as a way to adjust the topicality of nodes and explore the impact of user preferences on the probability of propagation.

**2.4 BAT algorithm**

The binary state network is the basic type of network and is widely used in the path architecture of various fields, including transmission [65, 66], network [67-69], data analysis [70], etc. Therefore, in recent years, research on binary state networks has been carried out and applied to all the above systems. With the growing scale of network, a more efficient and simple algorithm is needed to calculate the reliability of binary state networks to evaluate the performance and stability of networks.

The BAT algorithm was developed by Yeh in 2020 [33], which is an exhaustive method to find all possible state vectors and proposed a path-based hierarchical search algorithm to filter all connection vectors. This algorithm has been used in network reliability [33, 71], network resilience [72], wildfire propagation path prediction [73], etc. Compared with other path finding algorithms, such as Minimal Path algorithm (MP) or Minimal Cut algorithm (MC), the time complexity of BAT algorithm is the most efficient in reliability calculation of binary state network [33]. In other exhaustive algorithms, the experimental results confirm that the computer memory and efficiency required by the BAT algorithm are also better than the common depth-first search algorithm and breadth-first search algorithm [33].

The BAT algorithm uses simple binary addition to generate all possible state vectors. Since it is a binary state, the value of any coordinate in the state vector is 0 or 1. In terms of social network propagation, when the status is 1, it means that the linked page has been propagated; when the status is 0, it means that the linked page has not been propagated.

Suppose a binary state network graph $G(V, E)$ has $m$ coordinates and each edge has only two states, i.e., 0 or 1, $V$={0, 1, 2, …, $n$−1}, $X_k$ is the $k$th obtained state vector, and the $i$th coordinate $X(a_i)$ is the state of one of the edges $a_i$, where $a_i \in E$. The complete BAT algorithm is shown in Table 2 [33].

Table 2. BAT algorithm

| Input:    | $G(V, E)$ |
|---|---|
| Output:   | All possible non-repetitive state vectors. |

Step B0: Let SUM = 0, $k = 1$, $X_1 = X$, and $X$ is a zero vector with $m$ coordinates.

Step B1: Assume $i = m$.

Step B2: If $X(a_i) = 0$, then $X(a_i) = 1$, $k = k + 1$, $X_k = X$, SUM = SUM + 1, proceed to Step B4.

Step B3: If $i > 1$, let $X(a_i) = 0$, SUM = SUM – 1, $i = i – 1$, and go to Step B2.

Step B4: If SUM = $m$, stop. At this time $X_1$, $X_2$, ..., $X_k$ are all possible state vectors. Otherwise, proceed to Step B1.

As shown in Step B0, all coordinates start with a zero vector. The binary number of $X_k$ is updated by adding 1, as listed in Steps B1 to B3. The exhaustive process can be stopped by repeating the above steps until all coordinates are equal to 1, i.e., SUM= $m$.

The BAT algorithm proposes a more efficient, easy-to-understand, easy-to-program, and easy-to-modify path search algorithm for binary state networks. Therefore, we apply the BAT algorithm to the problem formulated in this study.

## 3. Research Methods

The main purpose of this study is to predict the propagation probability of social networks in different propagation ranges and then find out the key nodes in the network so that enables to achieve the maximum propagation benefits on social network through a small amount of investment costs. In this study, the spread of propagation paths in social networks is considered, and its final destination and number of destinations are unknown. Thus, it is difficult to estimate what might happen compared to regular network problems. This section focuses on the state presentation of network propagation and the application of algorithms, and establish a series of research methods.

### 3.1 Mathematics Notions

| $G(V,E)$ | Graph of binary state network. |
|---|---|
| $V$ | The total number of nodes included in the network graph, $V = \{0, 1, 2, …, n–1\}$ |
| $E$ | The total number of edges included in the network graph. |
| $i$ | Node, $i \in V$ |
| $V(i)$ | The set of adjacent points of node $i$ |
| $\mathrm{Deg}(i)$ | The total number of neighbors of node $i$ |
| $PR(i)$ | PageRank value of node $i$ |
| $S_k(i)$ | The $k$th state of node $i$, $k = 0, 1, \ldots, 2^{|\mathrm{Deg}(i)|}-1$ |
| $X$ | State vector of propagation path. |
| $X(i)$ | The state of node $i$. |
| $PR(S_k(i))$ | PageRank value of state $S_k(i)$. |
| $Pr(S_k(i))$ | The probability of occurrence of state $S_k(i)$. |
| $C(i)$ | Total number of state combinations for node $i$, $C(i) = 2^{|\mathrm{Deg}(i)|}$ |
| $PR_{max}(i)$ | The PageRank value of maximum propagation state for node $i$ |
| $N_{page}$ | At least the number of nodes to be propagated including the propagation source node. |
| $Pr(s, N_{page})=R$ | Probability of propagating to $N_{page}$ nodes starting from $s$ |
| $j$ | Newly propagated node. |
| $s$ | The source node of the propagation. |
| $T_l$ | The set of nodes that have propagated at state stage $l$ |
| $T^*$ | The set of nodes to be propagated at state stage $l$ |
| $P_l$ | Propagation probability for nodes that have already propagated ($T_l$) |
| $R_l$ | The newly added path probability at state stage $l$ |

### 3.2 Propagation status

The nodes status of social network is divided into two types: having been propagated and not being propagated. This section uses bridge network as an example, as shown in Figure 6. Each node $i$ belongs to the node set $V$ of the network graph, i.e. $i$

∈ $V$, and each node can be regarded as a user or page that is propagated in the social networks.

Node 0 connects node 1 and node 2, i.e. $Deg(i)=2$ and $V(i)=\{2,1\}$. The propagation path of node 0 includes propagation to ∅, {1}, {2}, and {1,2}, respectively, representing that information is propagated from node 0 but not propagated; information is propagated from node 0 and is not propagated to other nodes after being propagated to node 1; information is propagated from node 0 and is not propagated to other nodes after being propagated to node 2; information is propagated from node 0 and propagates to node 1 and node 2, and then stops propagating.

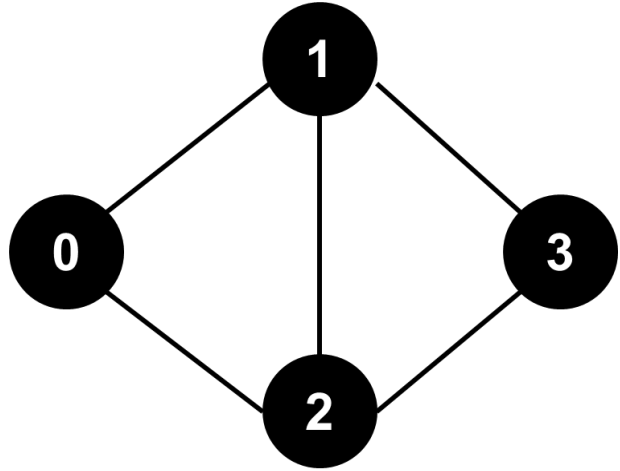

Figure 6. Network $G(V, E)$

The PageRank value of node $i$: $PR(i)$ is obtained by the PageRank algorithm in Section 2.3. Because the PageRank value represents the probability of the node appearing in the network, all PageRank values in the network add up to 1. Table 3 lists the total number of neighbors, the set of neighbors, and the PageRank value of node $i$ in the network in Figure 6.

Table 3. Basic information of network in Figure 6.

| $i$ | $Deg(i)$ | $V(i)$ | $PR(i)$ |
|---|---|---|---|
| 0 | 2 | {2,1} | 0.204787 |
| 1 | 3 | {3,2,0} | 0.295213 |
| 2 | 3 | {3,1,0} | 0.295213 |
| 3 | 2 | {2,1} | 0.204787 |

### 3.2.1 State labels and State vectors

Let $S_k(i)$ be the $k$th state of node $i$, which represents the different propagation states of nodes and also known as state labels. Because BAT algorithm representation are all binary and in order to successfully enumerate all node states without repetition, the binary state labels are converted to node state labels according to the rules in Table 4

[73].

<p style="text-align:center">Table 4. Transition rules of binary state label.</p>

| | |
|---|---|
| Rule 1: | In binary status labels, the number of bits ($m$) of the state value is equal to Deg($i$). |
| Rule 2: | In $V(i)$, node $i$ is in descending node order according to its label number. |
| Rule 3: | In the binary state label, the $m$th digit is equal to 0 or 1, 0 means that the $m$th node is not included in the state and 1 means that the $m$th node is included in the state. |

Taking the network $G(V, E)$ in Figure 6 as an example, enumerate all state vectors of node 0 by BAT algorithm. According to rule 1, the number of bits of the state value of node 0 is 2. According to rule 2, the $V(0)$ equals {2, 1} by descending arrangement. According to rule 3, the state labels exhaustively enumerated by BAT algorithm can be represented by binary state labels as 00, 01, 10, 11, respectively. That is, $S_0(0)= \varnothing$, $S_1(0)= \{1\}$, $S_2(0)= \{2\}$, $S_3(0)= \{1, 2\}$, hence, it can be seen that node 0 is propagated to $\varnothing$, {1}, {2}, and {1, 2}. Table 5 lists the node state labels and binary state labels of the example network.

<p style="text-align:center">Table 5. State label of network in Figure 6.</p>

| Node \ State | 0 | | 1 | | 2 | | 3 | |
|---|---|---|---|---|---|---|---|---|
| $V(i)$ | {2,1} | | {3,2,0} | | {3,1,0} | | {2,1} | |
| $S_k(i)$ | Node State Label | Binary State Label | Node State Label | Binary State Label | Node State Label | Binary State Label | Node State Label | Binary State Label |
| $S_0(i)$ | $\varnothing$ | [0 0] | $\varnothing$ | [0 0 0] | $\varnothing$ | [0 0 0] | $\varnothing$ | [0 0] |
| $S_1(i)$ | {1} | [0 1] | {0} | [0 0 1] | {0} | [0 0 1] | {1} | [0 1] |
| $S_2(i)$ | {2} | [1 0] | {2} | [0 1 0] | {1} | [0 1 0] | {2} | [1 0] |
| $S_3(i)$ | {2, 1} | [1 1] | {2, 0} | [0 1 1] | {1, 0} | [0 1 1] | {2, 1} | [1 1] |
| $S_4(i)$ | | | {3} | [1 0 0] | {3} | [1 0 0] | | |

| | | | | | |
|---|---|---|---|---|---|
| $S_5(i)$ | | {3, 0} | [1 0 1] | {3, 0} | [1 0 1] |
| $S_6(i)$ | | {3, 2} | [1 1 0] | {3, 1} | [1 1 0] |
| $S_7(i)$ | | {3, 2, 0} | [1 1 1] | {3, 1, 0} | [1 1 1] |

In order to record the propagated path of social network, this study uses $X$ as the state vector to record the node state label of propagation. The state vector is presented with a diagonal line to identify the propagation state, the right side of diagonal line is the propagation node and the left side of the diagonal line is the current state of propagation node. Taking $X = (2/1)$ of the network $G(V, E)$ in Figure 6 as an example, it represents state 2 of node1, which is transferred to node 2 that can be found from Table 5. If $X = (2/0, 4/2, 0/3)$, it means that the propagation starts from node 0, it propagates to node 2 after state 2 of node 0 occurs, then state 4 of node 2 occurs after passing to node 2 and is passed to node 3, and node 3 is in state 0 at this time indicating that propagation is stopped.

In addition to recording the propagated path in social networks, the source node and sink node can also be clearly known from the state vector. Exhausting the propagation paths in the form of state vectors can reduce the repetition of the propagation paths on the one hand, and have a comprehensive understanding of the propagation paths is on the other hand.

### 3.2.2 State probability and Propagation probability

In Section 2, it is mentioned that the social network is a scale-free network, that is, the Barabási-Albert model with a growth mode and a preferential connection mode. In preferential connection mode, the probability that a new node can be found to connect to node $i$ is proportional to the node's connectivity $k_i$ so the connection probability of web page is particularly important. Using the PageRank value introduced in Section 2.3 as probability, the $PR(S_k(i))$ is defined as the probability of $S_k(i)$, called state probability. When this node occurs state $S_k(i)$, then define $Pr(S_k(i))$ as the probability of occurrence of state $S_k(i)$. In other words, $Pr(S_k(i))$ can be regarded as the probability of node propagation. $PR(S_k(i))$ is to calculate the connection probability obtained by a node according to the PageRank values of all nodes in the state corresponding to node $i$ [73], as shown in Eq. (4).

$$PR(S_k(i)) = \sum_{|S_j(i)|=1 \text{ and } S_j(i) \subseteq S_k(i)} PR(S_j(i)) \qquad (4)$$

Because all possible state probabilities must sum to 1, the PageRank value is normalized as in Eq.(5).

$$Pr(S_k(i)) = \frac{PR(S_k(i))}{\sum_{k=0}^{2^{|Deg(i)|}-1} PR(S_k(i))} \qquad (5)$$

In all nodes, the probability of state 0 is non-transitive. From the preferential connection mode in the Barabási-Albert model, a newly added node can freely connect to any node in the network and the probability of connecting to two nodes is twice as high as connecting to one node [73], hence, the probability of state 0 is defined as Eq. (6).

$$PR(S_0(u)) = 0.5 \times \text{Min}\{ PR(i)，i \in V(u)\} \qquad (6)$$

Taking node 1 of the network in Figure 6 as an example, the state probability and propagation probability are shown in Table 6.

Table 6. State probabilities and propagation probabilities for node 1 of network in Figure 6.

| State | Propagation node | $PR(S_k(i))$ | $Pr(S_k(i))$ |
|---|---|---|---|
| $S_0(1)$ | ∅ | $0.5 \times \text{Min}\{PR(0)、PR(2)、PR(3)\}$ $=0.5 \times 0.204787 = 0.102394$ | 0.102394/2.921542= 0.0350478 |
| $S_1(1)$ | {0} | $PR(0) = 0.204787$ | 0.204787/2.921542= 0.0700956 |
| $S_2(1)$ | {2} | $PR(2) = 0.295213$ | 0.295213/2.921542= 0.101047 |
| $S_3(1)$ | {0, 2} | $PR(0) + PR(2) = 0.5$ | 0.5/2.921542= 0.171142 |
| $S_4(1)$ | {3} | $PR(3) = 0.204787$ | 0.204787/2.921542= 0.0700956 |
| $S_5(1)$ | {0, 3} | $PR(0) + PR(3) = 0.409574$ | 0.409574/2.921542= 0.140191 |
| $S_6(1)$ | {2, 3} | $PR(2) + PR(3) = 0.5$ | 0.5/2.921542= 0.171142 |
| $S_7(1)$ | {0,2,3} | $PR(0) + PR(2) + PR(3) = 0.704787$ | 0.704787/2.921542= 0.241238 |
| Sum | | 2.921542 | 1 |

According to the above method, the state probability and the propagation probability of network in Figure 6 are obtained as shown in Tables 7-8, respectively.

Table 7. State probability for network in Figure 6.

| $i$ | 0 | 1 | 2 | 3 |
|---|---|---|---|---|
| $V(i)$ | {2 1} | {3 2 0} | {3 1 0} | {2 1} |
| $PR(S_0(i))$ | 0.147606 | 0.102394 | 0.102394 | 0.147606 |
| $PR(S_1(i))$ | 0.295213 | 0.204787 | 0.204787 | 0.295213 |
| $PR(S_2(i))$ | 0.295213 | 0.295213 | 0.295213 | 0.295213 |
| $PR(S_3(i))$ | 0.590426 | 0.5 | 0.5 | 0.590426 |
| $PR(S_4(i))$ | | 0.204787 | 0.204787 | |
| $PR(S_5(i))$ | | 0.409574 | 0.409574 | |
| $PR(S_6(i))$ | | 0.5 | 0.5 | |
| $PR(S_7(i))$ | | 0.704787 | 0.704787 | |

Table 8. Propagation probability for network in Figure 6

| $i$ | 0 | 1 | 2 | 3 |
|---|---|---|---|---|
| $V(i)$ | {2 1} | {3 2 0} | {3 1 0} | {2 1} |
| $Pr(S_0(i))$ | 0.111111 | 0.035048 | 0.035048 | 0.111111 |
| $Pr(S_1(i))$ | 0.222222 | 0.070096 | 0.070096 | 0.222222 |
| $Pr(S_2(i))$ | 0.222222 | 0.101047 | 0.101047 | 0.222222 |
| $Pr(S_3(i))$ | 0.444444 | 0.171142 | 0.171142 | 0.444444 |
| $Pr(S_4(i))$ | | 0.070096 | 0.070096 | |
| $Pr(S_5(i))$ | | 0.140191 | 0.140191 | |
| $Pr(S_6(i))$ | | 0.171142 | 0.171142 | |
| $Pr(S_7(i))$ | | 0.241238 | 0.241238 | |

### 3.2.3 PageRank value of maximum propagation state

When a node reaches the maximum propagation state, it means that the node propagates to the subset of all possible nodes of $V(i)$ at the same time. Because the propagation state of each node is obtained by the BAT algorithm, it stops when all state vector coordinates are 1 during the exhaustive process so that the maximum propagation state of each node is the last state vector in the process of BAT algorithm. For example, the maximum state of node 0 is to propagate to node 1 and node 2 at the same time, that is, the binary state label is [1,1]. And the $C(i)$ is used to present the total number of state combinations. Because it is a binary network model, $C(i) = 2^{|Deg(i)|}$, and the value of $k$ in the state label $S_k(i)$ is encoded from 1 so that $k = 2^{|Deg(i)|}-1$ for the maximum propagation state.

The PageRank value of each node represents the importance of the corresponding page, and a page linked by many pages will have a higher PageRank value. When the node propagation state reaches the maximum state, that is, it propagates to all adjacent nodes at the same time. At this time, let $PR_{max}(i)$ be the PageRank value of maximum propagation state. Each node has a maximum propagation state. Taking the network in Figure 6 as an example, the maximum propagation state of node 1 is to propagate to node 0, node 2 and node 3 at the same time. At this time, the PageRank value of maximum propagation state of node 1 is the sum of the PageRank values of node 0, node 2 and node 3 [73], that is, $PR_{max}(1)$= 0.204787+0.295213+0.204787=0.704787. Table 9 shows the set of adjacent points, the total number of state combinations, the PageRank value and the PageRank value of maximum propagation state for each node of the network in Figure 6.

Table 9. Values related to the propagation state of network in Figure 6.

| $i$ | $V(i)$ | $C(i)$ | $PR(i)$ | $PR_{max}(i)$ |
|---|---|---|---|---|
| 0 | {2,1} | 4 | 0.204787 | 0.295213+0.295213=0.590426 |
| 1 | {3,2,0} | 8 | 0.295213 | 0.204787+0.295213+0.204787=0.704787 |
| 2 | {3,1,0} | 8 | 0.295213 | 0.204787+0.295213+0.204787=0.704787 |
| 3 | {2,1} | 4 | 0.204787 | 0.295213+0.295213=0.590426 |

## 3.3 A New BAT algorithm

The BAT algorithm can satisfy most of the networks with known conditions, such as the path exhaustion of network for one source node and one sink node, one source node with multiple sink nodes, or multiple source nodes to multiple sink nodes [71]. However, the propagation of social network is a source node and an unknown sink node. In addition to the unknown number of sink nodes, the node propagation state is also unknown. In order to predict the actual propagation situation of social network and realize path search by combining different states, the traditional BAT algorithm is modified so as to increase the suitability of the BAT algorithm to the practical problems discussed in this study. Assuming that information is propagated from node $s$, the purpose is to find all possible situations and the corresponding probabilities that the propagation of social network propagates to at least $N_{page}$ pages. Table 10 shows the process of the new BAT algorithm.

Table 10. The new BAT algorithm.

| | |
|---|---|
| Input: | $G(V, E)$, $V(i)$, $S_k(i)$, $s$, $N_{page}$ |
| Output: | $Pr(s, N_{page}) = R$ |
| Step A0: | Let $i = s$, $l = 0$, $R = 0$, $T_l = \{i\}$, $P_l = 1$, $X(i) = 1$, go to Step A1. |
| Step A1: | If $X(i) = 2^{|Deg(i)|}$, then node $i$ has exceeded the maximum propagation state, go to Step A8. |

Step A2: Let $T^* = \{j \mid j \in [V(i) - T_l]\}$.

Step A3: Let $T_{l+1} = T_l \cup T^*$. If $T^* = \varnothing$, then go to Step A6.

Step A4: When the number of nodes in $T^* \cup T_l$ is greater than or equal to $N_{page}$, then $R_l = P_l \times Pr(S_{X(i)}(i))$, $R = R + R_l$, and let $l = l+1$, $T_l = T_{l-1}$ and go to Step A7.

Step A5: $X(j) = 0$ for all $j \in T^*$, $P_l = P_{l-1} \times Pr(S_{X(i)}(i))$.

Step A6: Let $l = l+1$, if $i$ has the next node in $T_l$, update $i$ to be the next node and go to Step A1.

Step A7: Let $X(i) = X(i) + 1$, go to Step A1.

Step A8: If $i$ has a previous node in $T_l$, update $i$ to be the previous node and go to Step A7; if $i$ does not have a previous node in $T_l$, stop searching and $R$ is the final propagation probability at this time.

As shown in Step A0, the initial state vector includes the first node to start propagation, thus, its state starts from $X(i) = 1$. The other nodes that are not the initial propagation node must start from state 1 as shown in Step A5. At this time, Step A1 determines whether the node has reached the maximum state. If it reaches the maximum state, it needs to replace the node or stop searching as shown in Step A8. If the node does not reach the maximum state, then go to Step A2 and select the unconnected node in the adjacent node set as the node set to be propagated. It is determined by Step A3 whether the node set to be propagated is an empty set. If it is an empty set, it must go to Step A6 to update the node or go to Step A7 to add the node state. If it is judged by Step A3 that the node set to be propagated has newly added nodes, then go to Step A4 to judge whether it matches the number of $N_{page}$. If it matches the number of $N_{page}$, add its probability $R_l$ to the required $R$, then go to Step A7 to increase the state and continue to search for other propagation states. If the node has reached the maximum state and no other nodes can be updated, stop searching as shown in Step A8.

**4. Simulation experiment**

This study simulates the propagation in social network and presents the research results in Section 4. The proposed algorithms in this study are all coded in Python with version 3.6 and the simulation experiments are all executed on Spyder using a Windows 10 laptop with an Intel Core i5-8250U CPU(1.60 GHz，8 GB RAM).

**4.1 Model introduction**

NetworkX is a Python package that can randomly generate different types of networks and can also be used to analyze network structures, build network models, or draw networks. Using NetworkX to generate Barabási-Albert models with different number of nodes can ensure that the network model conforms to the power distribution and preferential connection mode [74], as shown in Figures 7-8.

A Barabási-Albert model with 8 nodes and 12 edges is randomly generated by NetworkX in Python as shown in Figure 7. Its degree distribution conforms to the power distribution as shown in Figure 8.

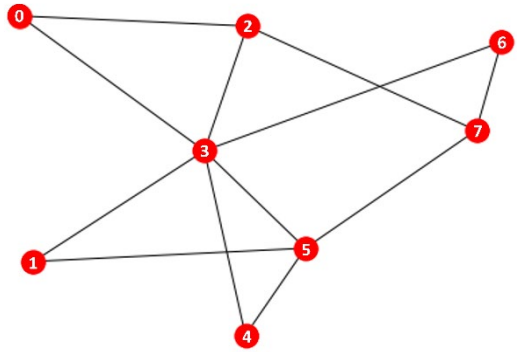

Figure 7. Network of Barabási-Albert model with 8 nodes and 12 edges

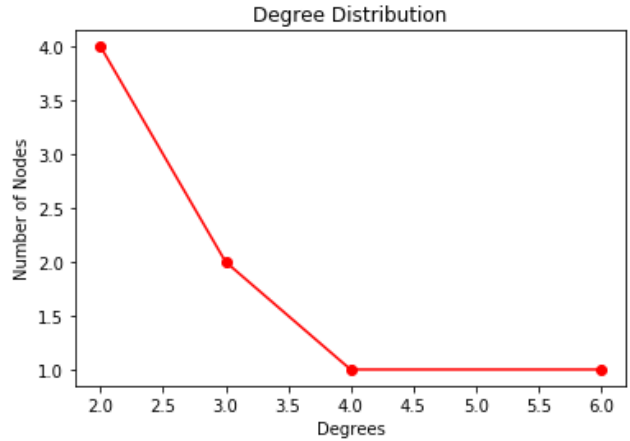

Figure 8. Degree distribution of network in Figure 7

Figure 7 shows the network graph used to simulate the scale-free network in this study, which contains 8 nodes and 12 edges where $i = 0, 1,…, 7$ and $i \in V$. In the experiment, we assume that the source node of propagation starts from $i = 0, 1, …, 7$, respectively, and consider all the propagation probability to $N_{page} = 1, 2, …, 8$. Therefore, this experiment contains a total of 64 experimental results.

Due to the influence of user preference on the propagation probability [25-27], the assigned weight of each node in the random walk varies according to the user preference of web page as shown in Section 2.3.2. For example, an advertiser plans to advertise a food. One of the nodes in the network of Figure 7 is a food-related website. At this time, the personalization vector of this node is $r_u = 1$ and the personalization vectors of the other nodes are $r_u = 0$. Thus, the model performs an experiment of assigning weights to three different nodes as follows.

Case 1: All nodes are assigned the same weight, that is, the traditional PageRank is used to calculate the page probability and the propagation probability is $Pr(s, N_{page})=R_1$.

Case 2: Set the node 3, which the total number of original adjacent nodes is the highest, as the preference node and use the personalized PageRank to calculate the probability of the page, that is, $r_3=1$, $r_0=r_1=r_2=r_4=r_5=r_6=r_7=0$. At this time, the propagation probability $Pr(s, N_{page})$ equals $R_2$.

Case 3: Set the node 1, which the original PageRank value is the lowest, as the preference node and use the personalized PageRank to calculate the probability of the page, that is, $r_1=1$, $r_0=r_2=r_3=r_4=r_5=r_6=r_7=0$. At this time, the propagation probability $Pr(s, N_{page})$ equals $R_3$.

In addition to analyzing and predicting the propagation probability of social network, this Section also changes theme of the node in disguise through the personalized vector. If the node is more in line with the user's preference, the allocation weight of the node is increased to interfere with the PageRank value. Table 11 presents the total number of adjacent nodes for node $i$, the set of adjacent nodes, the total number of state combinations, the PageRank value, the PageRank value of maximum propagation state, the personalized PageRank value, and the personalized PageRank value of maximum propagation state that are denoted as $Deg(i)$, $V(i)$, $C(i)$, $PR(i)$, $PR_{max}(i)$, $PPR(i)$, and $PPR_{max}(i)$, respectively.

Table 11. Basic information of the network graph in Figure 7

| Case | | | | Case 1 | | Case 2 ($r_3=1$) | | Case 3 ($r_1=1$) | |
|---|---|---|---|---|---|---|---|---|---|
| $i$ | $Deg(i)$ | $V(i)$ | $C(i)$ | $PR(i)$ | $PR_{max}(i)$ | $PPR(i)$ | $PPR_{max}(i)$ | $PPR(i)$ | $PPR_{max}(i)$ |
| 0 | 2 | {3 2} | 4 | 0.0879 | 0.3626 | 0.0795 | 0.4526 | 0.0569 | 0.3191 |
| 1 | 2 | {5 3} | 4 | 0.0867 | 0.3990 | 0.0789 | 0.4859 | 0.2226 | 0.4203 |
| 2 | 3 | {7 3 0} | 8 | 0.1254 | 0.4512 | 0.1088 | 0.5162 | 0.0824 | 0.3807 |
| 3 | 6 | {6 5 4 2 1 0} | 64 | 0.2372 | 0.6367 | 0.3438 | 0.5633 | 0.2367 | 0.6762 |
| 4 | 2 | {5 3} | 4 | 0.0867 | 0.3990 | 0.0789 | 0.4859 | 0.0726 | 0.4203 |
| 5 | 4 | {7 4 3 1} | 16 | 0.1618 | 0.5368 | 0.1421 | 0.5945 | 0.1836 | 0.6189 |
| 6 | 2 | {7 3} | 4 | 0.0881 | 0.3633 | 0.0750 | 0.4367 | 0.0582 | 0.3238 |
| 7 | 3 | {6 5 2} | 8 | 0.1261 | 0.3753 | 0.0929 | 0.3260 | 0.0871 | 0.3242 |

## 4.2 Results and discussion

Table 12 shows the probability of propagation to $N_{page}$ nodes and the running time of Case 1 with node $i$ as the propagation source node.

Table 12. Propagation probability results for the network in Figure 7

| $i$ | $N_{page}$ | $R_1$ runtime | $R_1$ | $R_2$ | $R_2-R_1$ | $R_3$ | $R_3-R_1$ |
|---|---|---|---|---|---|---|---|
| 0 | 1 | 0.0000 | 1.0000 | 1.0000 | 0.0000 | 1.0000 | 0.0000 |
|   | 2 | 0.0020 | 0.9204 | 0.9433 | 0.0229 | 0.9394 | 0.0190 |
|   | 3 | 0.0030 | 0.9071 | 0.9345 | 0.0274 | 0.9313 | 0.0242 |
|   | 4 | 0.0040 | 0.8923 | 0.9187 | 0.0265 | 0.9197 | 0.0274 |
|   | 5 | 0.0120 | 0.8681 | 0.8890 | 0.0209 | 0.8954 | 0.0273 |
|   | 6 | 0.0788 | 0.8163 | 0.8261 | 0.0098 | 0.8395 | 0.0233 |
|   | 7 | 0.6682 | 0.6992 | 0.6940 | -0.0052 | 0.7164 | 0.0172 |
|   | 8 | 4.7575 | 0.4056 | 0.3947 | -0.0109 | 0.4214 | 0.0159 |
| 1 | 1 | 0.0000 | 1.0000 | 1.0000 | 0.0000 | 1.0000 | 0.0000 |
|   | 2 | 0.0020 | 0.9079 | 0.9319 | 0.0239 | 0.9015 | -0.0064 |
|   | 3 | 0.0030 | 0.9007 | 0.9264 | 0.0257 | 0.8884 | -0.0123 |
|   | 4 | 0.0040 | 0.8878 | 0.9136 | 0.0259 | 0.8703 | -0.0174 |
|   | 5 | 0.0120 | 0.8536 | 0.8734 | 0.0198 | 0.8261 | -0.0274 |
|   | 6 | 0.0838 | 0.7997 | 0.8081 | 0.0085 | 0.7610 | -0.0386 |
|   | 7 | 0.7416 | 0.6704 | 0.6660 | -0.0044 | 0.6198 | -0.0506 |
|   | 8 | 4.6496 | 0.3942 | 0.3836 | -0.0106 | 0.3484 | -0.0458 |
| 2 | 1 | 0.0000 | 1.0000 | 1.0000 | 0.0000 | 1.0000 | 0.0000 |
|   | 2 | 0.0020 | 0.9762 | 0.9811 | 0.0049 | 0.9817 | 0.0054 |
|   | 3 | 0.0030 | 0.9563 | 0.9685 | 0.0122 | 0.9695 | 0.0132 |
|   | 4 | 0.0050 | 0.9401 | 0.9542 | 0.0141 | 0.9588 | 0.0187 |
|   | 5 | 0.0140 | 0.9171 | 0.9275 | 0.0104 | 0.9380 | 0.0209 |
|   | 6 | 0.1087 | 0.8644 | 0.8658 | 0.0013 | 0.8865 | 0.0221 |
|   | 7 | 0.8953 | 0.7145 | 0.7020 | -0.0125 | 0.7301 | 0.0156 |
|   | 8 | 5.5236 | 0.3756 | 0.3608 | -0.0148 | 0.3849 | 0.0093 |
| 3 | 1 | 0.0000 | 1.0000 | 1.0000 | 0.0000 | 1.0000 | 0.0000 |
|   | 2 | 0.0030 | 0.9979 | 0.9979 | 0.0000 | 0.9987 | 0.0008 |
|   | 3 | 0.0030 | 0.9901 | 0.9891 | -0.0009 | 0.9906 | 0.0005 |
|   | 4 | 0.0050 | 0.9674 | 0.9626 | -0.0048 | 0.9674 | 0.0000 |
|   | 5 | 0.0199 | 0.9196 | 0.9073 | -0.0123 | 0.9111 | -0.0085 |
|   | 6 | 0.1437 | 0.8383 | 0.8169 | -0.0214 | 0.8253 | -0.0131 |
|   | 7 | 1.0767 | 0.6683 | 0.6404 | -0.0279 | 0.6514 | -0.0169 |
|   | 8 | 6.1967 | 0.3471 | 0.3259 | -0.0213 | 0.3403 | -0.0068 |
| 4 | 1 | 0.0000 | 1.0000 | 1.0000 | 0.0000 | 1.0000 | 0.0000 |
| | 2 | 0.0020 | 0.9079 | 0.9319 | 0.0239 | 0.9015 | -0.0064 |
| | 3 | 0.0030 | 0.9007 | 0.9264 | 0.0257 | 0.8961 | -0.0047 |
| | 4 | 0.0030 | 0.8878 | 0.9136 | 0.0259 | 0.8814 | -0.0064 |
| | 5 | 0.0130 | 0.8536 | 0.8734 | 0.0198 | 0.8388 | -0.0148 |
| | 6 | 0.0911 | 0.7997 | 0.8081 | 0.0085 | 0.7786 | -0.0210 |
| | 7 | 0.7126 | 0.6704 | 0.6660 | -0.0044 | 0.6458 | -0.0246 |
| | 8 | 4.5909 | 0.3942 | 0.3836 | -0.0106 | 0.3881 | -0.0061 |
| 5 | 1 | 0.0000 | 1.0000 | 1.0000 | 0.0000 | 1.0000 | 0.0000 |
| | 2 | 0.0020 | 0.9900 | 0.9918 | 0.0018 | 0.9927 | 0.0027 |
| | 3 | 0.0030 | 0.9745 | 0.9817 | 0.0072 | 0.9720 | -0.0026 |
| | 4 | 0.0060 | 0.9582 | 0.9683 | 0.0102 | 0.9518 | -0.0063 |
| | 5 | 0.0209 | 0.9233 | 0.9298 | 0.0064 | 0.9068 | -0.0166 |
| | 6 | 0.1337 | 0.8561 | 0.8520 | -0.0041 | 0.8304 | -0.0257 |
| | 7 | 0.9995 | 0.6894 | 0.6736 | -0.0158 | 0.6599 | -0.0294 |
| | 8 | 5.9696 | 0.3574 | 0.3413 | -0.0161 | 0.3447 | -0.0127 |
| 6 | 1 | 0.0000 | 1.0000 | 1.0000 | 0.0000 | 1.0000 | 0.0000 |
| | 2 | 0.0030 | 0.9201 | 0.9495 | 0.0293 | 0.9370 | 0.0168 |
| | 3 | 0.0030 | 0.9046 | 0.9387 | 0.0341 | 0.9273 | 0.0228 |
| | 4 | 0.0030 | 0.8940 | 0.9288 | 0.0348 | 0.9186 | 0.0246 |
| | 5 | 0.0100 | 0.8739 | 0.9049 | 0.0310 | 0.8991 | 0.0252 |
| | 6 | 0.0760 | 0.8211 | 0.8423 | 0.0211 | 0.8427 | 0.0215 |
| | 7 | 0.6580 | 0.7012 | 0.7086 | 0.0074 | 0.7161 | 0.0149 |
| | 8 | 4.4182 | 0.4036 | 0.3967 | -0.0069 | 0.4186 | 0.0150 |
| 7 | 1 | 0.0000 | 1.0000 | 1.0000 | 0.0000 | 1.0000 | 0.0000 |
| | 2 | 0.0020 | 0.9715 | 0.9720 | 0.0005 | 0.9781 | 0.0065 |
| | 3 | 0.0030 | 0.9463 | 0.9555 | 0.0092 | 0.9617 | 0.0154 |
| | 4 | 0.0049 | 0.9333 | 0.9486 | 0.0153 | 0.9508 | 0.0176 |
| | 5 | 0.0180 | 0.9207 | 0.9382 | 0.0176 | 0.9396 | 0.0189 |
| | 6 | 0.1271 | 0.8725 | 0.8871 | 0.0146 | 0.8930 | 0.0205 |
| | 7 | 0.8932 | 0.7200 | 0.7245 | 0.0045 | 0.7335 | 0.0135 |
| | 8 | 5.7668 | 0.3773 | 0.3721 | -0.0052 | 0.3857 | 0.0084 |

From Table 12, it can be found that the propagation probability is 100% when $N_{page}$ is 1 because the propagation source node in social network starts to propagate itself whether in Case 1, Case 2 or Case 3. That is $\Pr(i, N_{page}=1)=1$. When the propagation range increases, that is, $N_{page}$ increases, the propagation probability of each node decreases, which means that propagation probability of reaching more nodes is lower and the propagation probability is negatively correlated with the propagation range.

**4.2.1 Neighbors and propagation probability of node $i$**

From the results of $R_1$ and $R_2$ in Table 12, if the adjacent nodes of two different nodes are the same, the probability of propagating to $N_{page}$ nodes is the same, that is, $Pr(i, N_{page})= Pr(j, N_{page})$ if $V(i)=V(j)$. For example, in Figure 7, the adjacent nodes of node 1 and node 4 are node 5 and node 3, that is $V(1)=V(4)$. At this time, the probability of propagating to $N_{page} =1, 2, …, 8$ is all the same that are 1.0000, 0.9079, 0.9007, 0.8878, 0.8536, 0.7997, 0.6704 and 0.3942, respectively. However, in $R_3$, because the personalized vector changes the assigned weight of node 1, the PageRank value of node 1 is increased resulting in different propagation probabilities even if the adjacent nodes of node 1 and node 4 are the same. Therefore, it can be judged that if the adjacent connected nodes are the same and the PageRank values of the nodes are the same, the probability of the final propagating to $N_{page}$ nodes is the same although the propagation paths are different. And the propagation probability is different if the PageRank values are different due to the intervention of the personalized vector.

The total number of adjacent nodes also affects the propagation probability. It can be seen from Table 12 that when the number of adjacent nodes is propagated to a smaller number of $N_{page}$, the node with more total adjacent nodes has a higher propagation probability; on the contrary, when it propagates to a larger number of $N_{page}$, the node with smaller total adjacent nodes has a higher propagation probability. For example, the total number of adjacent nodes of node 3 and node 5 are the first and the second respectively, which are the nodes with the higher total number of adjacent nodes in Figure 7. Whether in Case 1, Case 2 or Case 3, when $N_{page} =0, 1, …, 4$, node 3 and node 5 have higher propagation probabilities than other nodes, especially node 3 with the largest number of adjacent nodes. But when $N_{page} =8$, node 3 has the lowest propagation probability compared to other nodes. On the contrary, the total number of adjacent nodes of node 0, node 1, node 4, and node 6 is 2, which is the node with the lowest total number of adjacent nodes in Figure 7. When $N_{page} =0, 1, …, 7$, node 0 has the lowest propagation probability compared to other nodes. But when $N_{page} =8$, the propagation probability of node 0 compared with other nodes: $R_1$=0.4056 and $R_3$=0.4214 ranks first, and $R_2$=0.3947 ranks second. Hence, the node with a larger total number of adjacent nodes can be selected as the source node when need to propagate a small range. However, in the final stage of propagation, nodes with fewer number of adjacent nodes have a higher probability of propagating to more pages.

**4.2.2 PageRank value and propagation probability of maximum propagation state**

The PageRank value of the maximum propagation state means that it will be propagated to all adjacent nodes at the same time if the node is the source node, that is, the maximum propagation state is reached. And the PageRank value of reaching the maximum propagation state is also related to the propagation probability. Table 13 shows the ranking of PageRank value of maximum propagation state relative to other nodes in the three cases. Combining the information analysis of Table 12 and Table 13, the following results are obtained.

In terms of a smaller propagation range, the node with a larger PageRank value of maximum propagation state has a higher propagation probability. For the larger propagation range, the node with a smaller PageRank value of maximum propagation state has a higher propagation probability. For example, the PageRank values of maximum propagation state of node 3 and node 5 are ranked higher than other nodes, and the propagation probability corresponding to $N_{page}$=1, 2, …, 4 are all among the top. However, the propagation probability of node 3 including $R1$, $R2$, and $R3$ are all the lowest when $N_{page}$=8. On the other hand, nodes 0, 6 and 7 with lower PageRank values of maximum propagation state have higher propagation probability when the propagation range is large.

It can be seen from the above results that to prevent the propagation of malicious messages, it is recommended to prioritize the protection of nodes with a higher PageRank value of maximum propagation state at the beginning of the propagation of messages, which can effectively and quickly prevent the propagation of messages.

Table 13. Ranking for PageRank value of maximum propagation state

| $i$ | Case 1 $PR_{max}(i)$ | Ranking | Case 2 $PPR_{max}(i)$ | Ranking | Case 3 $PPR_{max}(i)$ | Ranking |
|---|---|---|---|---|---|---|
| 0 | 0.3626 | 7 | 0.4526 | 6 | 0.3191 | 8 |
| 1 | 0.3990 | 4 | 0.4859 | 4 | 0.4203 | 3 |
| 2 | 0.4512 | 3 | 0.5162 | 3 | 0.3807 | 5 |
| 3 | 0.6367 | 1 | 0.5633 | 2 | 0.6762 | 1 |
| 4 | 0.3990 | 4 | 0.4859 | 4 | 0.4203 | 3 |
| 5 | 0.5368 | 2 | 0.5945 | 1 | 0.6189 | 2 |
| 6 | 0.3633 | 6 | 0.4367 | 7 | 0.3238 | 7 |
| 7 | 0.3753 | 8 | 0.3260 | 8 | 0.3242 | 6 |

In order to verify that the conclusions in Subsections 4.2.1 and 4.2.2 are not mode coincidences, several scale-free networks with different numbers of nodes and edges are built with NetworkX. The models and propagation probabilities are attached in Appendix Table1. Judging from the propagation probability of other models: if the set of adjacent nodes of a node is the same, the probability of propagating to $N_{page}$ nodes is the same; when the propagation range is small, nodes with a higher total number of adjacent nodes and a larger PageRank value of maximum propagation state have a higher propagation probability, which are consistent with the above conclusion.

**4.2.3 Personalization vectors and propagation probability**

In the propagation process of social network, the authority, professionalism, trust and influence of propagation nodes are all related to user preferences but most influence studies of social network largely ignore this part [27]. In this study, the personalized vector in the personalized PageRank algorithm is used to intervene in the random viewers in the PageRank algorithm to prefer a specific node in the random walk process so as to increase the distribution weight of the node to represent the user's preference node or relevance of information propagation.

In Case 2, we set random viewers to prefer a specific node 3 in the random walk process, which node 3 is originally the node with the highest number of adjacent nodes and the highest PageRank value. From Table 11, it can be found that when the personalized vector $r_3=1$ of node 3, $PPR(3)=0.3438$ is higher than the original $PR(3)=0.2372$. However, in the $R_2-R_1$ column in Table 12, it can be found that most of the propagation probabilities increase, while the propagation probability of node 3 decreases. The reason is because node 3 connects most of the nodes resulting that it affects the PageRank value of maximum propagation state of other nodes also increases when the PageRank value of node 3 increases, as shown in Table 14.

Table 14. In Case 2, the PageRank value of maximum propagation state increases and decreases.

| $i$ | $PR_{max}(i)$ | $PPR_{max}(i)$ | Increase/Decrease | Rank change |
|---|---|---|---|---|
| 0 | 0.3626 | 0.4526 | 0.0900 | 7→6 |
| 1 | 0.3990 | 0.4859 | 0.0869 | 4→4 |
| 2 | 0.4512 | 0.5162 | 0.0650 | 3→3 |
| 3 | 0.6367 | 0.5633 | -0.0734 | 1→2 |
| 4 | 0.3990 | 0.4859 | 0.0869 | 4→4 |
| 5 | 0.5368 | 0.5945 | 0.0577 | 2→1 |
| 6 | 0.3633 | 0.4367 | 0.0734 | 6→7 |
| 7 | 0.3753 | 0.3260 | -0.0494 | 8→8 |

In Case 3, we set random viewers to prefer a specific node 1 in the random walk process, which node 1 is originally the node with the lowest number of adjacent nodes and the lowest PageRank value. When the PageRank value of node 1 increases, it affects the PageRank value of maximum propagation state of other nodes. However, because there are not many nodes connected to node 1, the PageRank value of maximum propagation state does not increase much. As shown in Table 15, the number of nodes with increased propagation probability is small. From the $R_3-R_1$ column in Table 12, it can be found that most of the propagation probabilities decrease, which means that increasing the personalized vector of nodes with fewer connections has no obvious effect on the propagation probability of nodes.

Table 15. In Case 3, the PageRank value of maximum propagation state increases and decreases.

| $i$ | $PR_{max}(i)$ | $PPR_{max}(i)$ | Increase/Decrease | Rank change |
|---|---|---|---|---|
| 0 | 0.3626 | 0.3191 | -0.0436 | 7→8 |
| 1 | 0.3990 | 0.4203 | 0.0213 | 4→3 |
| 2 | 0.4512 | 0.3807 | -0.0705 | 3→5 |
| 3 | 0.6367 | 0.6762 | 0.0395 | 1→1 |
| 4 | 0.3990 | 0.4203 | 0.0213 | 4→3 |
| 5 | 0.5368 | 0.6189 | 0.0821 | 2→2 |
| 6 | 0.3633 | 0.3238 | -0.0395 | 6→7 |
| 7 | 0.3753 | 0.3242 | -0.0511 | 8→6 |

When we bias the random viewers to the nodes with more adjacent nodes in the random walk process, the propagation probability of more nodes can be increased. However, increasing the distribution weight of nodes with fewer adjacent nodes does not have much influence. Therefore, it is assumed that an advertiser plans to place an advertisement and there are two nodes with the same theme. If a node with a higher number of adjacent nodes is selected, it relatively increases the probability of propagating more pages.

## 5. Conclusions

The contribution of this research is to propose a new method, which is novel, simple and easy to apply, to calculate the propagation probability of social network propagating to the rest of the web pages. After confirming that the social network is a scale-free network, the Barabási- Albert model is used to build a model. Then, the PageRank algorithm and the personalized PageRank algorithm are adopted to calculate the occurrence probability and propagation probability of each node. After enumerating all possible propagation states of each node by the BAT algorithm, the new BAT algorithm is used to calculate the propagation probability under different propagation ranges.

The new BAT algorithm can calculate the predicted value of $Pr(s, N_{page})$ for all nodes $i$ and $N_{page}$, that is, the propagation probability of social network, as shown in Table 12. In addition, the personalized vector in the personalized PageRank algorithm is used to intervene the preferences of random viewers in the random walk process so that the prediction model is more in line with the actual user situation.

The proposed scale-free network model, new BAT algorithm, PageRank value of maximum propagation state, and personalized vector concept can be extended to other network problems in the future.

**Acknowledgments**

This research was supported in part by the Ministry of Science and Technology, R.O.C (MOST 107-2221-E-007-072-MY3, MOST 110-2221-E-007-107-MY3, MOST 109-2221-E-424-002 and MOST 110-2511-H-130-002).

and function using NetworkX. *Los Alamos National Lab.(LANL)*, Los Alamos, NM (United States), 2008.

# Appendix

Table 1. Propagation probability results for different numbers of nodes and edges

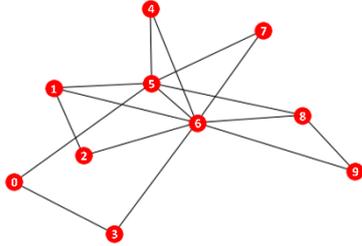

Model 1: 10 nodes and 16 edges.

| Npage  | 1 | 2 | 3 | 4 | 5 | 6 | 7 | 8 | 9 | 10 |
|---|---|---|---|---|---|---|---|---|---|---|
| Node 0 | 1 | 0.9343 | 0.9112 | 0.9078 | 0.9045 | 0.8976 | 0.8692 | 0.7730 | 0.5548 | 0.2411 |
| Node 1 | 1 | 0.9830 | 0.9754 | 0.9725 | 0.9660 | 0.9486 | 0.8938 | 0.7505 | 0.4958 | 0.2067 |
| Node 2 | 1 | 0.9335 | 0.9264 | 0.9239 | 0.9177 | 0.9035 | 0.8610 | 0.7411 | 0.5065 | 0.2279 |
| Node 3 | 1 | 0.9458 | 0.9240 | 0.9219 | 0.9168 | 0.9066 | 0.8751 | 0.7766 | 0.5564 | 0.2408 |
| Node 4 | 1 | 0.9027 | 0.9014 | 0.8983 | 0.8925 | 0.8769 | 0.8259 | 0.7003 | 0.4657 | 0.2137 |
| Node 5 | 1 | 0.9984 | 0.9940 | 0.9889 | 0.9823 | 0.9604 | 0.8893 | 0.7256 | 0.4599 | 0.1879 |
| Node 6 | 1 | 0.9996 | 0.9979 | 0.9924 | 0.9812 | 0.9540 | 0.8807 | 0.7184 | 0.4560 | 0.1867 |
| Node 7 | 1 | 0.9027 | 0.9014 | 0.8983 | 0.8925 | 0.8769 | 0.8259 | 0.7003 | 0.4657 | 0.2137 |
| Node 8 | 1 | 0.9830 | 0.9754 | 0.9725 | 0.9660 | 0.9486 | 0.8938 | 0.7505 | 0.4958 | 0.2067 |
| Node 9 | 1 | 0.9335 | 0.9264 | 0.9239 | 0.9177 | 0.9035 | 0.8610 | 0.7411 | 0.5065 | 0.2279 |

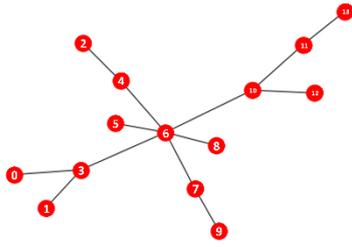

Model 2: 14 nodes and 13 edges.

| Npage | 1 | 2 | 3 | 4 | 5 | 6 | 7 | 8 | 9 | 10 |
|---|---|---|---|---|---|---|---|---|---|---|
| Node 0 | 1 | 0.6667 | 0.6309 | 0.5570 | 0.5424 | 0.5007 | 0.4254 | 0.3264 | 0.2233 | 0.1329 |
| Node 1 | 1 | 0.6667 | 0.6309 | 0.5570 | 0.5424 | 0.5007 | 0.4254 | 0.3264 | 0.2233 | 0.1329 |
| Node 2 | 1 | 0.6667 | 0.5836 | 0.5797 | 0.5588 | 0.5089 | 0.4273 | 0.3247 | 0.2206 | 0.1307 |
| Node 3 | 1 | 0.9821 | 0.9090 | 0.8256 | 0.7852 | 0.6995 | 0.5702 | 0.4188 | 0.2729 | 0.1536 |
| Node 4 | 1 | 0.9585 | 0.8727 | 0.8553 | 0.8043 | 0.7080 | 0.5713 | 0.4164 | 0.2699 | 0.1524 |
| Node 5 | 1 | 0.6667 | 0.6639 | 0.6464 | 0.6002 | 0.5191 | 0.4127 | 0.2950 | 0.1886 | 0.1050 |
| Node 6 | 1 | 0.9986 | 0.9858 | 0.9400 | 0.8463 | 0.7061 | 0.5376 | 0.3681 | 0.2240 | 0.1183 |
| Node 7 | 1 | 0.9585 | 0.8727 | 0.8553 | 0.8043 | 0.7080 | 0.5713 | 0.4164 | 0.2699 | 0.1524 |

| | | | | | | | | | |
|---|---|---|---|---|---|---|---|---|---|
| Node 8 | 1 | 0.6667 | 0.6639 | 0.6464 | 0.6002 | 0.5191 | 0.4127 | 0.2950 | 0.1886 | 0.1050 |
| Node 9 | 1 | 0.6667 | 0.5836 | 0.5797 | 0.5588 | 0.5089 | 0.4273 | 0.3247 | 0.2206 | 0.1307 |
| Node 10 | 1 | 0.9844 | 0.9280 | 0.8463 | 0.7596 | 0.6886 | 0.5729 | 0.4260 | 0.2762 | 0.1521 |
| Node 11 | 1 | 0.9325 | 0.7726 | 0.6960 | 0.6306 | 0.6004 | 0.5329 | 0.4259 | 0.2974 | 0.1762 |
| Node 12 | 1 | 0.6667 | 0.6354 | 0.5934 | 0.5215 | 0.4884 | 0.4243 | 0.3321 | 0.2278 | 0.1332 |
| Node 13 | 1 | 0.6667 | 0.5317 | 0.4917 | 0.4250 | 0.4139 | 0.3811 | 0.3190 | 0.2346 | 0.1472 |

| Npage | 11 | 12 | 13 | 14 |
|---|---|---|---|---|
| Node 0 | 0.0662 | 0.0260 | 0.0071 | 0.0011 |
| Node 1 | 0.0662 | 0.0260 | 0.0071 | 0.0011 |
| Node 2 | 0.0664 | 0.0272 | 0.0080 | 0.0012 |
| Node 3 | 0.0716 | 0.0260 | 0.0065 | 0.0009 |
| Node 4 | 0.0730 | 0.0278 | 0.0074 | 0.0010 |
| Node 5 | 0.0497 | 0.0189 | 0.0052 | 0.0008 |
| Node 6 | 0.0527 | 0.0186 | 0.0047 | 0.0006 |
| Node 7 | 0.0730 | 0.0278 | 0.0074 | 0.0010 |
| Node 8 | 0.0497 | 0.0189 | 0.0052 | 0.0008 |
| Node 9 | 0.0664 | 0.0272 | 0.0080 | 0.0012 |
| Node 10 | 0.0688 | 0.0242 | 0.0061 | 0.0008 |
| Node 11 | 0.0852 | 0.0318 | 0.0083 | 0.0011 |
| Node 12 | 0.0645 | 0.0244 | 0.0067 | 0.0010 |
| Node 13 | 0.0757 | 0.0302 | 0.0086 | 0.0013 |

Model 3: 20 nodes and 19 edges.

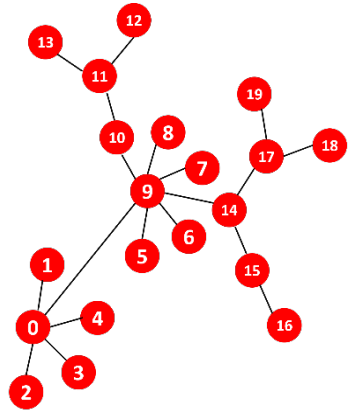

| Npage | 1 | 2 | 3 | 4 | 5 | 6 | 7 | 8 | 9 | 10 |
|---|---|---|---|---|---|---|---|---|---|---|
| Node 0 | 1 | 0.9968 | 0.9708 | 0.8913 | 0.8025 | 0.7439 | 0.6819 | 0.5957 | 0.4970 | 0.3929 |
| Node 1 | 1 | 0.6667 | 0.6602 | 0.6213 | 0.5539 | 0.5108 | 0.4788 | 0.4269 | 0.3633 | 0.2949 |
| Node 2 | 1 | 0.6667 | 0.6602 | 0.6213 | 0.5539 | 0.5108 | 0.4788 | 0.4269 | 0.3633 | 0.2949 |
| Node 3 | 1 | 0.6667 | 0.6602 | 0.6213 | 0.5539 | 0.5108 | 0.4788 | 0.4269 | 0.3633 | 0.2949 |
| Node 4 | 1 | 0.6667 | 0.6602 | 0.6213 | 0.5539 | 0.5108 | 0.4788 | 0.4269 | 0.3633 | 0.2949 |

| | | | | | | | | | |
|---|---|---|---|---|---|---|---|---|---|
| Node 5 | 1 | 0.6667 | 0.6655 | 0.6557 | 0.6300 | 0.5915 | 0.5347 | 0.4554 | 0.3687 | 0.2867 |
| Node 6 | 1 | 0.6667 | 0.6655 | 0.6557 | 0.6300 | 0.5915 | 0.5347 | 0.4554 | 0.3687 | 0.2867 |
| Node 7 | 1 | 0.6667 | 0.6655 | 0.6557 | 0.6300 | 0.5915 | 0.5347 | 0.4554 | 0.3687 | 0.2867 |
| Node 8 | 1 | 0.6667 | 0.6655 | 0.6557 | 0.6300 | 0.5915 | 0.5347 | 0.4554 | 0.3687 | 0.2867 |
| Node 9 | 1 | 0.9994 | 0.9931 | 0.9675 | 0.9194 | 0.8486 | 0.7475 | 0.6228 | 0.4955 | 0.3760 |
| Node 10 | 1 | 0.9241 | 0.9016 | 0.8218 | 0.7460 | 0.7123 | 0.6589 | 0.5815 | 0.4880 | 0.3881 |
| Node 11 | 1 | 0.9670 | 0.8099 | 0.5975 | 0.5400 | 0.5293 | 0.5038 | 0.4627 | 0.4077 | 0.3379 |
| Node 12 | 1 | 0.6667 | 0.6006 | 0.4251 | 0.3614 | 0.3578 | 0.3454 | 0.3217 | 0.2890 | 0.2464 |
| Node 13 | 1 | 0.6667 | 0.6006 | 0.4251 | 0.3614 | 0.3578 | 0.3454 | 0.3217 | 0.2890 | 0.2464 |
| Node 14 | 1 | 0.9778 | 0.9492 | 0.8799 | 0.8124 | 0.7379 | 0.6683 | 0.5995 | 0.5108 | 0.4093 |
| Node 15 | 1 | 0.9313 | 0.7719 | 0.7313 | 0.6871 | 0.6155 | 0.5613 | 0.5201 | 0.4603 | 0.3834 |
| Node 16 | 1 | 0.6667 | 0.5292 | 0.4939 | 0.4784 | 0.4291 | 0.3836 | 0.3609 | 0.3266 | 0.2788 |
| Node 17 | 1 | 0.9726 | 0.8513 | 0.7005 | 0.6422 | 0.5848 | 0.5423 | 0.5056 | 0.4512 | 0.3805 |
| Node 18 | 1 | 0.6667 | 0.6119 | 0.4835 | 0.4451 | 0.4052 | 0.3698 | 0.3503 | 0.3191 | 0.2761 |
| Node 19 | 1 | 0.6667 | 0.6119 | 0.4835 | 0.4451 | 0.4052 | 0.3698 | 0.3503 | 0.3191 | 0.2761 |
| $N_{page}$ | 11 | 12 | 13 | 14 | 15 | 16 | 17 | 18 | 19 | 20 |
| Node 0 | 0.2897 | 0.1976 | 0.1248 | 0.0726 | 0.0380 | 0.0172 | 0.0064 | 0.0018 | 0.0003 | 2.77E-05 |
| Node 1 | 0.2244 | 0.1576 | 0.1023 | 0.0615 | 0.0336 | 0.0160 | 0.0063 | 0.0019 | 0.0004 | 3.51E-05 |
| Node 2 | 0.2244 | 0.1576 | 0.1023 | 0.0615 | 0.0336 | 0.0160 | 0.0063 | 0.0019 | 0.0004 | 3.51E-05 |
| Node 3 | 0.2244 | 0.1576 | 0.1023 | 0.0615 | 0.0336 | 0.0160 | 0.0063 | 0.0019 | 0.0004 | 3.51E-05 |
| Node 4 | 0.2244 | 0.1576 | 0.1023 | 0.0615 | 0.0336 | 0.0160 | 0.0063 | 0.0019 | 0.0004 | 3.51E-05 |
| Node 5 | 0.2101 | 0.1425 | 0.0895 | 0.0522 | 0.0279 | 0.0133 | 0.0053 | 0.0016 | 0.0003 | 3.05E-05 |
| Node 6 | 0.2101 | 0.1425 | 0.0895 | 0.0522 | 0.0279 | 0.0133 | 0.0053 | 0.0016 | 0.0003 | 3.05E-05 |
| Node 7 | 0.2101 | 0.1425 | 0.0895 | 0.0522 | 0.0279 | 0.0133 | 0.0053 | 0.0016 | 0.0003 | 3.05E-05 |
| Node 8 | 0.2101 | 0.1425 | 0.0895 | 0.0522 | 0.0279 | 0.0133 | 0.0053 | 0.0016 | 0.0003 | 3.05E-05 |
| Node 9 | 0.2672 | 0.1760 | 0.1076 | 0.0609 | 0.0314 | 0.0142 | 0.0053 | 0.0015 | 0.0003 | 2.38E-05 |
| Node 10 | 0.2893 | 0.2014 | 0.1313 | 0.0790 | 0.0426 | 0.0197 | 0.0074 | 0.0021 | 0.0004 | 3.33E-05 |
| Node 11 | 0.2596 | 0.1866 | 0.1268 | 0.0802 | 0.0453 | 0.0217 | 0.0082 | 0.0023 | 0.0004 | 3.59E-05 |
| Node 12 | 0.1943 | 0.1422 | 0.0985 | 0.0643 | 0.0379 | 0.0191 | 0.0077 | 0.0023 | 0.0004 | 4.14E-05 |
| Node 13 | 0.1943 | 0.1422 | 0.0985 | 0.0643 | 0.0379 | 0.0191 | 0.0077 | 0.0023 | 0.0004 | 4.14E-05 |
| Node 14 | 0.3056 | 0.2110 | 0.1336 | 0.0767 | 0.0395 | 0.0178 | 0.0066 | 0.0018 | 0.0003 | 2.95E-05 |
| Node 15 | 0.2989 | 0.2164 | 0.1431 | 0.0857 | 0.0464 | 0.0222 | 0.0087 | 0.0025 | 0.0005 | 4.14E-05 |
| Node 16 | 0.2227 | 0.1660 | 0.1134 | 0.0699 | 0.0390 | 0.0195 | 0.0081 | 0.0025 | 0.0005 | 4.7E-05 |
| Node 17 | 0.2987 | 0.2157 | 0.1422 | 0.0854 | 0.0460 | 0.0215 | 0.0081 | 0.0023 | 0.0004 | 3.52E-05 |
| Node 18 | 0.2235 | 0.1666 | 0.1133 | 0.0702 | 0.0393 | 0.0193 | 0.0077 | 0.0023 | 0.0004 | 4.17E-05 |
| Node 19 | 0.2235 | 0.1666 | 0.1133 | 0.0702 | 0.0393 | 0.0193 | 0.0077 | 0.0023 | 0.0004 | 4.17E-05 |